\documentclass[fleqn,usenatbib]{mnras}

\usepackage{newtxtext,newtxmath,bm}
\usepackage{natbib} 

\usepackage[T1]{fontenc}

\DeclareRobustCommand{\VAN}[3]{#2}
\let\VANthebibliography\thebibliography
\def\thebibliography{\DeclareRobustCommand{\VAN}[3]{##3}\VANthebibliography}

\usepackage{graphicx}	
\usepackage{amsmath}	

\newcommand{\emunit}{\mathrm{cm}^{-6} \ \mathrm{pc}}
\newcommand{\sgmunit}{\mathrm{M}_{\odot} \mathrm{pc}^{-2}}
\newcommand{\cc}{\mathrm{cm}^{-3}}

\newcommand{\msun}{\mathrm{M}_{\odot}}

\newcommand{\nel}{n_\mathrm{e}}
\newcommand{\tff}{t_\mathrm{ff}}
\newcommand{\hii }{H\,{\sc ii} }
\newcommand{\oi}{O\,{\sc i}}
\newcommand{\cii}{C\,{\sc ii}}
\newcommand{\oiii}{O\,{\sc iii}}

\newcommand{\neav}{\langle n_\mathrm{e} \rangle}
\newcommand{\Td}{\bar{T}_\mathrm{d}}

\defcitealias{FY21}{FY21}
\defcitealias{Hunt09}{HH09}

\title[Observational signatures of forming YMCs]{Observational signatures of forming young massive clusters: \\
continuum emission from dense \hii regions}

\author[M. Inoguchi et al.]{
Mutsuko Inoguchi,$^{1}$\thanks{E-mail: mutsuko.inoguchi@gmail.com}
Takashi Hosokawa,$^{2}$\thanks{E-mail:
hosokawa@tap.scphys.kyoto-u.ac.jp}
Hajime Fukushima,$^{3}$,
Kei E. I. Tanaka,$^{4}$, 
\newauthor
Hidenobu Yajima,$^{3}$,
Shin Mineshige,$^{1}$
\\
$^{1}$Department of Astronomy, Graduate School of Science, Kyoto University, Sakyo, Kyoto 606-8502, Japan\\
$^{2}$Department of Physics, Graduate School of Science, Kyoto University, Sakyo, Kyoto 606-8502, Japan\\
$^{3}$Center for Computational Sciences, University of Tsukuba, Ten-nodai, 1-1-1 Tsukuba, Ibaraki 305-8577, Japan\\
$^{4}$Department of Earth and Planetary Sciences, Tokyo Institute of Technology, Meguro, Tokyo, 152-8551, Japan
}

\date{Accepted XXX. Received YYY; in original form ZZZ}

\pubyear{2023}

\begin{document}
\label{firstpage}
\pagerange{\pageref{firstpage}--\pageref{lastpage}}
\maketitle

\begin{abstract}
Young massive clusters (YMCs) are the most massive star clusters forming in nearby galaxies and are thought to be a young analogue to the globular clusters. Understanding the formation process of YMCs leads to looking into very efficient star formation in high-redshift galaxies suggested by recent JWST observations. We investigate possible observational signatures of their formation stage, particularly when the mass of a cluster is increasing via accretion from a natal molecular cloud. To this end, we study the broad-band continuum emission from ionized gas and dust enshrouding YMCs, whose formation is followed by recent radiation-hydrodynamics simulations. We perform post-process radiative transfer calculations using simulation snapshots and find characteristic spectral features at radio and far-infrared frequencies. We show that a striking feature is long-lasting, strong free-free emission from a $\sim$ 10~pc-scale \hii region with a large emission measure of $\gtrsim 10^7 \emunit$, corresponding to the mean electron density of $\gtrsim 10^3~\cc$. There is a turnover feature below $\sim$ 10~GHz, a signature of the optically thick free-free emission, often found in Galactic ultra-compact \hii regions. These features come from the peculiar YMC formation process, where the cluster's gravity effectively traps photoionized gas for a long duration and enables continuous star formation within the cluster. Such large and dense \hii regions show distinct distribution on the density-size diagram, apart from the standard sequence of Galactic \hii regions. This is consistent with the observational trend inferred for extragalactic \hii regions associated with YMCs.     
\end{abstract}



\section{Introduction}
\label{sec:intro}


Young massive clusters (YMCs), also known as superstar clusters, are the most massive star clusters forming in the present-day universe \citep[][]{PZ2010,Longmore14}. 
Their typical mass ($\gtrsim 10^4~\msun$) and density ($\gtrsim 10^3~\msun$ pc$^{-3}$) are much larger than those of open clusters (OCs), typical star clusters in our Galaxy \citep{Beck15}. YMCs are often found in nearby starburst and interacting galaxies \citep[e.g.][]{Whitmore1993, Whitmore1999, Whitmore00}, and they are considered a young analogue to the globular clusters. Therefore, revealing the YMC formation processes in the nearby galaxies leads to glimpsing the star formation in the early universe. Some latest JWST observations resolve individual YMCs in distant galaxies even at redshifts $z \simeq 4-6$ \citep[e.g.][]{Vanzella2022a,Vanzella2022b}, suggesting that the YMCs should contribute larger fractions of the star formation activities in such galaxies. 


Star clusters in their early formation stage are called ``embedded clusters" \citep{Lada03}. Previous optical and infrared (IR) observations report candidates of YMCs in the embedded stage in nearby galaxies \citep[e.g.][]{Gorjian2001, Beck2002, Galliano2005}.  
The cold gas and dust associated with these candidates are confirmed by recent high-resolution observations using Atacama Large Millimeter/sub-millimeter Array (ALMA) \citep[][]{Johnson2015,Turner2017,Leroy2018,Finn2019}. More recently, the latest JWST observations report discoveries of new populations of YMCs, many of which seem to be in an early embedded stage \citep[][]{Whitmore23}. 
Recent ALMA observations by \citet{He22} report possible candidates of individual $\sim 10$~pc-scale \hii regions associated with embedded YMCs in the Antennae galaxy.  
Theoretical models suggest that YMCs and progenitors of globular clusters often form in highly pressurised environments \citep{Elmegreen1997,Kruijssen15}, and some observations point to cloud-cloud collisions as a key physical process \citep{Tsuge2021b,Tsuge2021a}. However, the exact mechanism of the YMC formation, including the evolution of a cluster under radiative feedback from high-mass cluster-member stars \citep{Krumholz19}, remains uncertain.


Recently, \citet{FY21} (FY21 hereafter) studied conditions required for the YMC formation, systematically performing a suite of radiation-hydrodynamic (RHD) simulations of the cluster formation and cloud destruction by the stellar radiative feedback \citep[see also, e.g.][]{Dale12, Kim18, He2019, Dobbs2020, Grudic2021, Menon2022}. As a result of examining cases with different parameters, such as the initial cloud mass, size, metallicity, etc., they found the critical gas surface density above which the YMC formation occurs, $\Sigma \simeq 200~\msun {\rm pc}^{-2}$.
Recent observations report that molecular clouds with such high surface densities are ubiquitously found in distant galaxies at redshifts $z \sim 1$ \citep{Dessauges-Zavadsky23}.
In the cases of the YMC formation, a star cluster rapidly becomes massive enough to gravitationally trap the hot photoionized gas, before the entire cloud is blown away by an expanding \hii bubble. The cluster mass continues to increase even after the emergence of the \hii bubble for such a case \citep{Bressert12}, resulting in a high star-formation efficiency (SFE), or the mass ratio of the cluster to the natal molecular cloud.


\citetalias{FY21} suggest characteristic features of the formation stage of the YMC, i.e., the stage when the cluster mass increases through accretion from the natal cloud. The gravity from the growing massive cluster keeps the \hii bubble compact and dense around it for a long duration \citep{Keto2003}. We aim to derive observational signatures of such a characteristic evolutionary stage. To this end, we conduct various post-process radiation transfer calculations using simulation snapshots. 
In this paper, we particularly investigate the continuum spectra expected for the YMC formation stage. We study the free-free radio spectrum from dense \hii regions associated with forming YMCs, with which the mean density of the photoionized gas has been observationally inferred \citep[e.g.][]{Garay1999,Kim01}. 
Intriguingly, \citet{Hunt09} (hereafter HH09) demonstrate that \hii regions found in Blue Compact Dwarf galaxies obey the density-size relation apart from that for Galactic \hii regions \citep[see also][]{Gilbert07}. 
They show that such extragalactic \hii regions are distinctively denser than their Galactic counterparts with similar sizes, suggesting a different population of \hii regions in extreme starburst environments. 
We show that our post-process calculations for OC- and YMC-forming cases well explain such observational trends regarding the density-size relations.


We organize the rest of the paper as follows. In~Section~\ref{sec:method}, we briefly review our numerical methods on the radiation hydrodynamic simulations by \citetalias{FY21}. We also describe the method for post-process radiation transfer calculations to give the continuum spectra based on the simulation data. In Section~\ref{ssec:FY21}, we compare the two representative cases of the OC and YMC formation, which we mainly consider throughout the paper. In Section~\ref{ssec:cont}, we consider the evolution of the continuum spectra from radio to IR wavelengths, with which we consider possible observational signatures of the YMC formation.
Finally, Sections~\ref{sec:discussion} and \ref{sec:conclusion} provide discussion and concluding remarks.

\section{Methods}
\label{sec:method}

\subsection{Radiation-hydrodynamics simulations}
\label{ssec:method_rhd}

Here we briefly describe the method of three-dimensional (3D) RHD simulations of cluster formation and cloud destruction \citepalias[see also][]{FY21}. They use a modified version of the adaptive mesh refinement (AMR) code \textsc{sfumato} \citep{Matumoto2007, Matsumoto15}, for which the M1 method \citep[e.g.][]{Levermore84} is implemented to handle the radiative transfer (dubbed \textsc{sfumato-m1}). They adopt the reduced speed of light approximation with $\tilde{c} = 3 \times 10^{-4} c$, where $c$ is the light speed \citep{Rosdahl2013}.
Photoionization of atoms, photodissociation of molecules, and radiative heating of gas and dust around a light source are solved by considering the transfer of extreme ultraviolet (EUV; $13.6~{\rm eV}<h\nu$), Lyman-Werner (LW; $11.2~{\rm eV} < h \nu < 13.6~{\rm eV}$), far-ultraviolet (FUV; $6~{\rm eV} < h \nu < 13.6 ~{\rm eV}$), and infrared photon (IR) photons \citep[e.g., ][]{HI06}. The chemistry solver is based on the scheme of \citet{Sugimura2020}, and it is extended to include the network of \citet{NL97} for CO formation. The chemical network also includes O$^0$, O$^+$, and O$^{2+}$, whose abundances are solved with the same procedure as in \citet{Fukushima20a}.


We insert sink particles representing a small star cluster when the density exceeds a threshold value and other conditions are satisfied \citep{Federrath2010}.
We assign photon emissivity to each sink particle in the following two ways. One is the same as in \citetalias{FY21}, where the luminosity and spectrum are given by taking the averages of the stellar isochrone of \citet{Chen2015} and the Chabrier initial mass function \citep[IMF,][]{C03}. The other is stochastic sampling, where the total cluster mass is distributed into stellar mass bins in a probability-weighted manner based on the IMF \citep{Fukushima2022}.

\subsection{Cases examined}

\begin{table}
\begin{center}
\caption{Cases considered}
\begin{tabular}{lccccc}\hline \hline
		Model & $M_{\rm cl}$ & $\Sigma_{\rm cl}$  & $R_{\rm cl}$ & $n_0$ & $\tff$ \\ 
		      & (M$_{\odot}$) &  (M$_{\odot}/$pc$^2$) & (pc) &  (cm$^{-3}$) & (Myr)  \\
		\hline
		M6R28st (YMC) & $10^6$ & 400 & 28.2 & 309 & 2.5 \\
		M6R56st (OC)  & $10^6$ & 100 & 56.4 & 38.6 & 7.0 \\ 
        M6R10 (YMC)  & $10^6$ &  3200 & 10.0 &  7000  & 0.52   \\
        M5R20 (OC)   & $10^5$ & 80  & 20.0 & 87 & 4.7  \\
        M6R60 (OC)   & $10^6$ & 88 & 60.0 & 32 & 7.7 \\
  \hline
\end{tabular}
\label{tab:simpar}
\end{center}
\end{table}
\begin{figure*}
	\includegraphics[width=140mm]{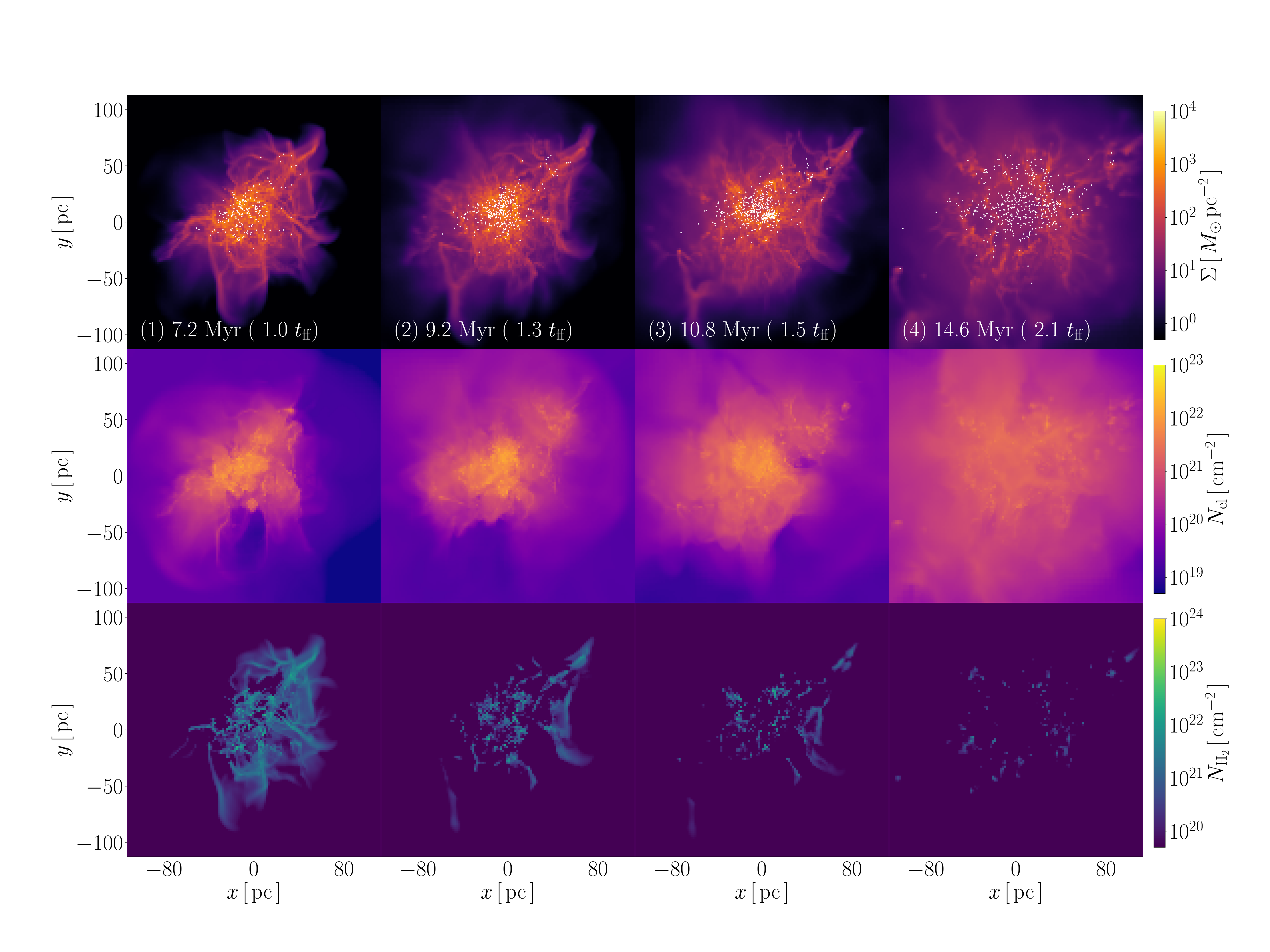}
\caption{Birth of a star cluster and destruction of a cloud simulated in the representative case of the OC formation, M6R56st. The columns of panels display snapshots at the different epochs, (1) $t= 7.2$~Myr, (2) 9.2~Myr, (3) 10.8~Myr, and (4) 14.6~Myr from left to right. The top, middle, and bottom rows display the distributions of the total gas surface density $\Sigma$, column density of electrons $N_\mathrm{el}$, and that of hydrogen molecules $N_{\mathrm{H}_2}$ measured along the line of sight, $z$-axis. The white dots represent the distributions of the star (cluster) particles.  
}
\label{fig:snap_sigma100m6}
\end{figure*}
\begin{figure*}
\includegraphics[width=140mm]{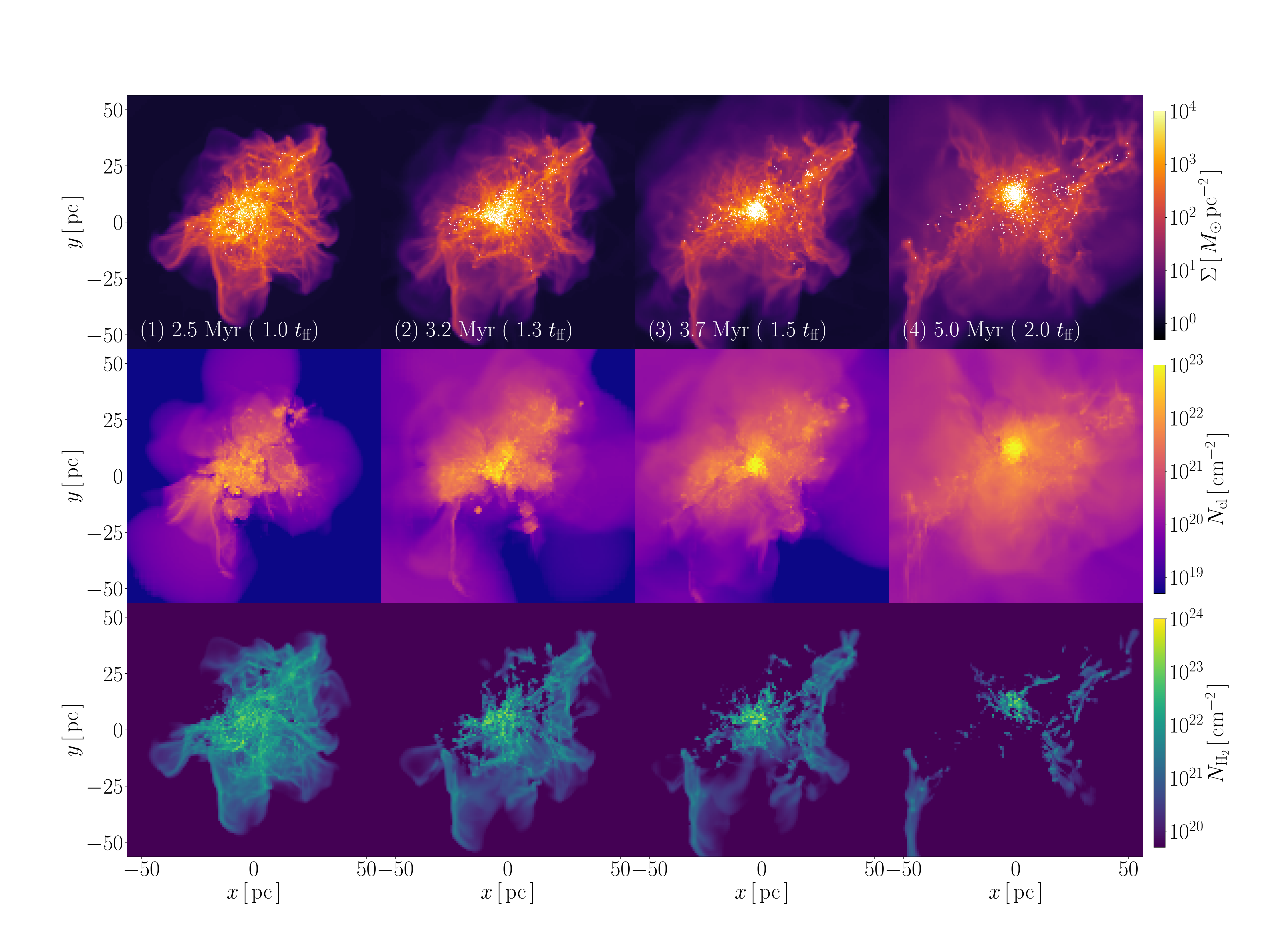}
\caption{The same as Fig.~\ref{fig:snap_sigma100m6} but for the representative case of the YMC formation, M6R28st. The columns of panels correspond to the different epochs of (1) $t= 2.5$~Myr, (2) 3.2~Myr, (3) 3.7~Myr, and (4) 5.0~Myr from left to right, the same times as in Fig.~\ref{fig:snap_sigma100m6} if normalized by the initial free-fall timescale $\tff$.}
\label{fig:snap_sigma400m6}
\end{figure*}
\begin{figure}
	\includegraphics[width=\columnwidth]{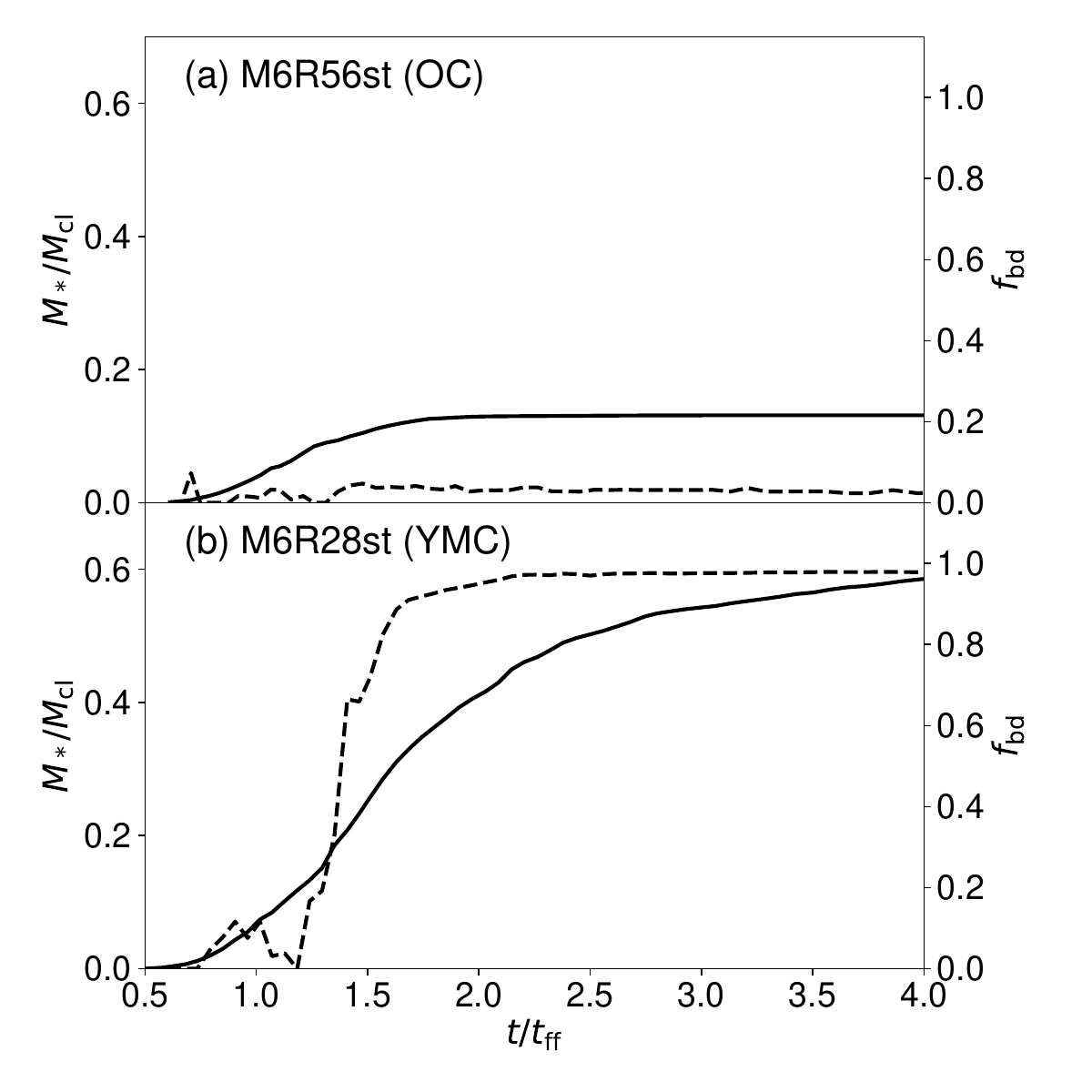}
\caption{Star formation histories against the elapsed time normalized by the free-fall timescale for the representative cases of the OC and YMC formation, M6R56st and M6R28st (panels a and b). In each panel, the solid and dashed lines represent the cluster mass normalized by the initial cloud mass $M_*/M_\mathrm{cl}$ and mass fraction of gravitationally bound particles $f_\mathrm{bd}$, respectively.}
    \label{fig:mass_history}
\end{figure}

Table~\ref{tab:simpar} summarizes the cases considered. Each simulation run starts from a homogeneous cloud with the mass $M_\mathrm{cl}$, surface density $\Sigma_\mathrm{cl}$, and radius $R_\mathrm{cl}$. We also assume that a turbulent velocity field fills the cloud. 
Throughout the paper, we particularly focus on two representative cases, M6R28st and M6R56st, for which the cloud mass is the same as $M_{\rm cl} = 10^6~\msun$. Case M6R28st starts from a relatively compact cloud with $\Sigma_{\rm cl} = 400~\sgmunit$, or with the radius $R_\mathrm{cl} \simeq 28$~pc. The free-fall timescale for the initial cloud is $\tff \simeq$ 2.5~Myr. 
The other case M6R56st starts from a cloud with $\Sigma_{\rm cl} = 100~\sgmunit$, or with the radius $R_\mathrm{cl} \simeq 56$~pc. The corresponding free-fall timescale is $\tff \simeq$ 7~Myr. 
As shown in Section~\ref{ssec:FY21} below, M6R28st and M6R56st represent the typical cases of yielding YMC-like and OC-like clusters, respectively. 
We newly perform these simulation runs using the stochastic stellar sampling developed in \citet{Fukushima2022} (Section~\ref{ssec:method_rhd}) as indicated by the label "st". \citet{Kim16} show that the stochasticity of the stellar population impacts the emissivity only when the total stellar mass is smaller than $10^4~\msun$ \citep[see also,][]{Fukushima2022}.
In our work, the cluster mass is much higher than this value for all the cases examined, and these cases provide the essentially same results as in \citetalias{FY21}.


Apart from the above cases, we consider three additional cases M6R10, M5R20, and M6R60, identical to those studied in \citetalias{FY21}, where the model names have extensions of "Z0A1". For example, case M6R10 corresponds to M6R10Z0A1 in \citetalias{FY21}.

\subsection{Post-process continuum radiative transfer calculations}


For post-process radiative transfer calculations, we convert the original AMR data of a simulation run into a data cube composed of $(128)^3$ cells.\footnote{
We also performed an experimental post-process radiation transfer calculation using a high-resolution data cube with $(256)^3$ cells, which only encloses a small central part of the original computational domain. The resulting cell size is 1/5.5 times smaller than the original size used for the default calculations. The resulting spectrum deviates from the corresponding default result only by a factor of less than 2.
} 
These cells are homogeneously distributed throughout the computational domain.
We map all the physical quantities on the original AMR grids to the Cartesian grids by linear interpolation. We solve the transfer equation at a given frequency, which ranges from $\nu_{\rm min} = 0.1$~GHz to $\nu_{\rm max} = 10^5$~GHz, along a line of sight chosen as an axis of the data cube,
\begin{equation}
\frac{d I_\nu}{ds} = - (\kappa_{\nu,{\rm ff}} + \kappa_{\nu,{\rm d}}) I_\nu + \kappa_{\nu,{\rm ff}} B_\nu(T) + \kappa_{\nu,{\rm d}} B_\nu(T_{\rm d}) ,
\end{equation}
where $I_\nu$ is the intensity, $\kappa_{\nu,{\rm ff}}$ the free-free opacity, $\kappa_{\nu,{\rm d}}$ the dust opacity, $B_\nu$ the Planck function, and $T$ and $T_{\rm d}$ are the gas and dust temperatures. The free-free opacity is written as 
\begin{equation}
\kappa_{\nu,{\rm ff}} = \frac{j_{\nu, {\rm ff}}}{B_\nu (T)} ,
\end{equation}
where $j_{\nu,{\rm ff}}$ is the free-free emission coefficient, which is given by 
\begin{eqnarray}
j_{\nu,{\rm ff}} &\simeq& 5.4 \times 10^{-41} {\rm erg} {\rm s}^{-1} {\rm cm}^{-3} {\rm Hz}^{-1} {\rm rad}^{-1} \\ \nonumber
& & \times \ \ T_4^{-1/2} n_e n_p \exp\left( - \frac{h \nu }{ k_{\rm B} T} \right) g_{\rm ff} ,
\end{eqnarray}
where $T_4 \equiv (T/10^4~{\rm K})$, $n_e$ and $n_p$ the electron and proton number densities, $h$ the Planck constant, $k_{\rm B}$ the Boltzmann constant, $g_{\rm ff}$ the Gaunt factor \citep[e.g.][]{Rybicki1986,Draine11a}. Regarding the dust opacity $\kappa_{\nu, {\rm d}}$, we use
\begin{equation}
\kappa_{\nu,\rm d} = 0.1 \left( \frac{\lambda}{300 \mu{\rm m}} \right)^{-2} {\rm cm}^2 {\rm g}^{-1},
\label{eq:dkappa}
\end{equation}
where $\lambda$ is the wavelength. 
The functional form of Eq.~\eqref{eq:dkappa} is valid for the far-IR range, and it has been typically assumed for the {\it Herschel} Infrared Galactic plane survey \citep[Hi-GAL, e.g.][]{Molinari10,Elia13}, which covers $70~\micron \lesssim \lambda \lesssim 500~\micron$. In our RHD simulations, the energy balance between the IR dust thermal emission, dust absorption of UV and IR photons, and gas-dust collisional energy exchange determines the local dust temperature $T_{\rm d}$. We use the Planck mean opacity given by \citet{Laor1993} for the monochromatic IR transfer.

\section{Results}
\label{sec:results}

\subsection{Bimodal evolution: OC and YMC formation}
\label{ssec:FY21}

Here we briefly review the evolution for cases M6R56st and M6R28st, representatives of the OC and YMC formations observed in our RHD simulations. 
\citetalias{FY21} also provide similar descriptions in more detail.


Fig.~\ref{fig:snap_sigma100m6} shows the time evolution of case M6R56st, where the surface density of the cloud is relatively low, $\Sigma_{\rm cl} = 100~\sgmunit$. In the early phase, the initial turbulent motion controls the gas dynamics and induces filamentary structures along which the star formation occurs.
An \hii region appears around the star cluster by the epoch of $t \simeq \tff$. The natal cloud gradually disperses owing to the expanding \hii region during $\tff \lesssim t \lesssim 2 \tff$. We see that the distribution of star cluster particles extends during that period. Most hydrogen molecules are destroyed by stellar FUV radiation \citep{Inoguchi2020}. The star cluster is only surrounded by photoionized gas at the final snapshot of $t \simeq 2 \tff$. 


Fig.~\ref{fig:snap_sigma400m6} shows the evolution of case M6R28st, where the initial cloud surface density is higher than the above as $\Sigma_{\rm cl} = 400~\sgmunit$.
While the basic evolution looks similar to that presented in Fig.~\ref{fig:snap_sigma100m6}, a denser and more compact star cluster finally appears in this case. As studied in \citetalias{FY21}, the gravitational force from the star cluster overwhelms the thermal pressure gradient caused by the \hii regions. The electron density near the cluster centre remains much higher than in case M6R56st. We derive observational signatures of such large and dense \hii regions in Section~\ref{ssec:cont} below. 


Fig.~\ref{fig:mass_history} presents the time evolution of the ratio of the cluster mass $M_*$ to the initial cloud mass $M_*/M_\mathrm{cl}$ and the bound fraction of the star cluster $f_\mathrm{bd}$, the mass fraction of star-cluster particles that are gravitationally bound. 
Both cases show a common evolution that $M_*/M_\mathrm{cl}$ reaches 0.1 at $t \simeq 1.3~\tff$.
However, the subsequent evolution is very different among these models. Whereas $M_*/M_\mathrm{cl}$ saturates at $\simeq 0.13$ at $t \simeq 1.5~\tff$ for case M6R56st, it continuously increases to reach 0.6 until $t \simeq 4~\tff$ for case M6R28st. Early theoretical studies generally predict that the bound fraction of the star cluster is a sensitive function of the SFE; high $f_\mathrm{fd}$ is achieved with the SFE exceeding a few $\times$ 10~\% after the dispersal of a natal cloud \citep[e.g.,][]{Adams2000, Baumgardt2007, Shukirgaliyev2017}. 
Similarly, \citetalias{FY21} show that the evolution when $M_*/M_\mathrm{cl} \sim 0.1$ determines the subsequent fate, whether it leads to the YMC formation. 
In the former case of M6R56st (see also Fig.~\ref{fig:snap_sigma100m6}), $M_*/M_\mathrm{cl}$ is relatively small at $t \simeq 1.3~\tff$, so that the bound fraction remains less than 0.1 throughout the simulation. The UV radiative feedback effectively suppresses the star formation no later than $t \simeq 1.5~\tff$. In the latter case of M6R28st (see also Fig.~\ref{fig:snap_sigma400m6}), $M_*/M_\mathrm{cl}$ becomes relatively large before the stellar UV feedback becomes effective, i.e., before the \hii bubble sweeps a large part of the original cloud. The bound fraction exceeds 0.9 at $t \simeq 1.5~\tff$, and the gravitational potential around the star cluster becomes deep enough to capture the ambient gas.


As mentioned in Section~\ref{sec:intro}, \citetalias{FY21} show that different initial cloud properties lead to the bimodal evolution overviewed for the above two cases M6R56st and M6R28st. Therefore, we particularly study the observational signatures of these cases, supposing that M6R56st and M6R28st demonstrate the typical YMC and OC formation processes, respectively.

\subsection{Continuum emission from ionized gas and dust around forming clusters}
\label{ssec:cont}

\subsubsection{Continuum spectra consisting of free-free emission and dust thermal emission}

\begin{figure}
	\includegraphics[width=\columnwidth]{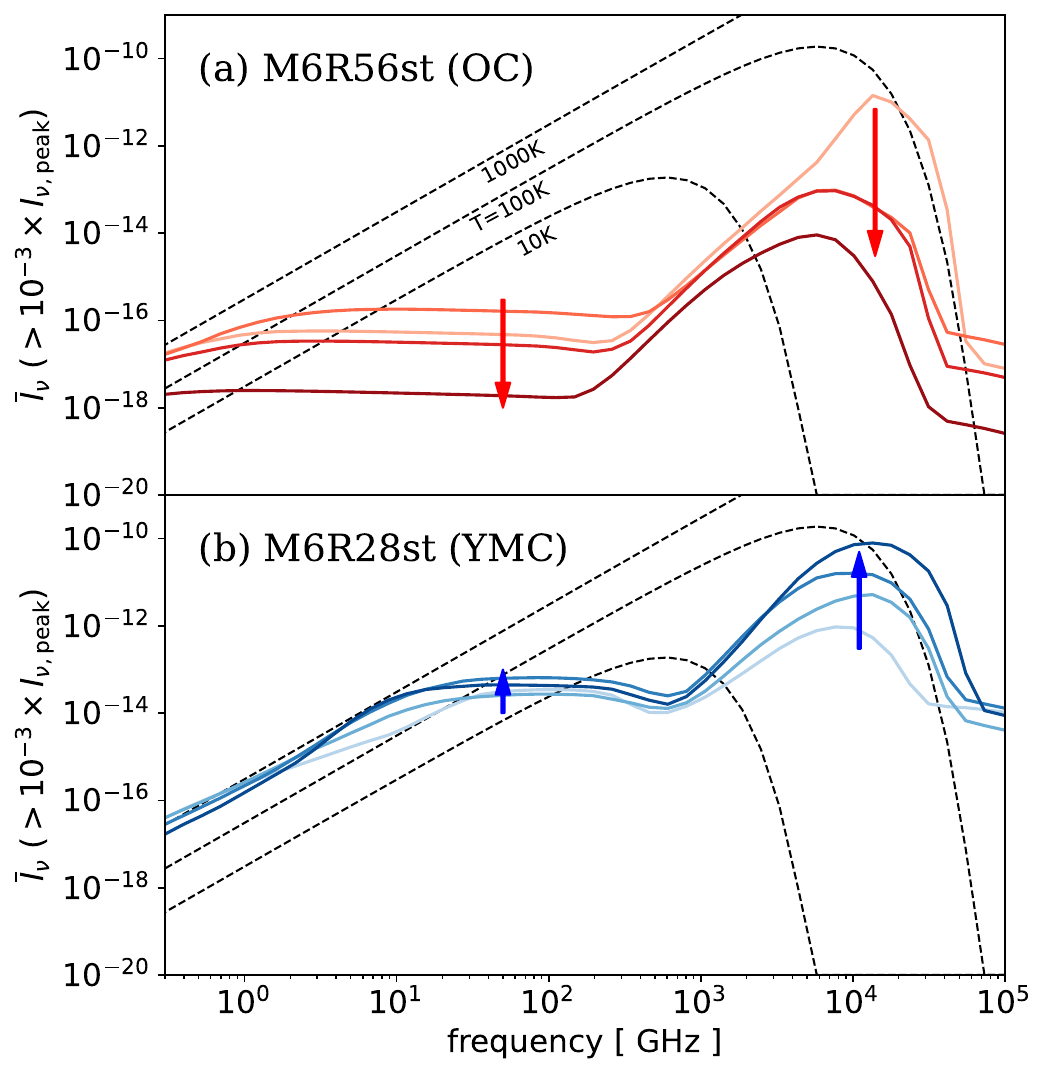}
    \caption{
Time evolution of the continuum spectrum emitted from a central part of cluster-forming regions. Panels (a) and (b) represent the cases of M6R56st and M6R28st, where the YMC and OC formation eventually occur. The vertical axis represents the intensity averaged over a part where the intensity is higher than 0.1~\% of the peak value at a given frequency. In both panels, the darker line colors represent the snapshots in the later stages of $t=1.0$, 1.3, 1.5, and 2.1 (2.0) $t_\mathrm{ff}$,
the same snapshots as in Fig.~\ref{fig:snap_sigma100m6} (Fig.~\ref{fig:snap_sigma400m6}). 
The direction of the vertical arrows denotes how the evolution proceeds. The thin dashed curves represent the reference spectrum of the Planck function $B_\nu (T_\mathrm{eff})$ with the effective temperatures $T_\mathrm{eff} = 1000$~K, 100~K, and 10~K.
}
\label{fig:contsed}
\end{figure}
\begin{figure}
	\includegraphics[width=\columnwidth]{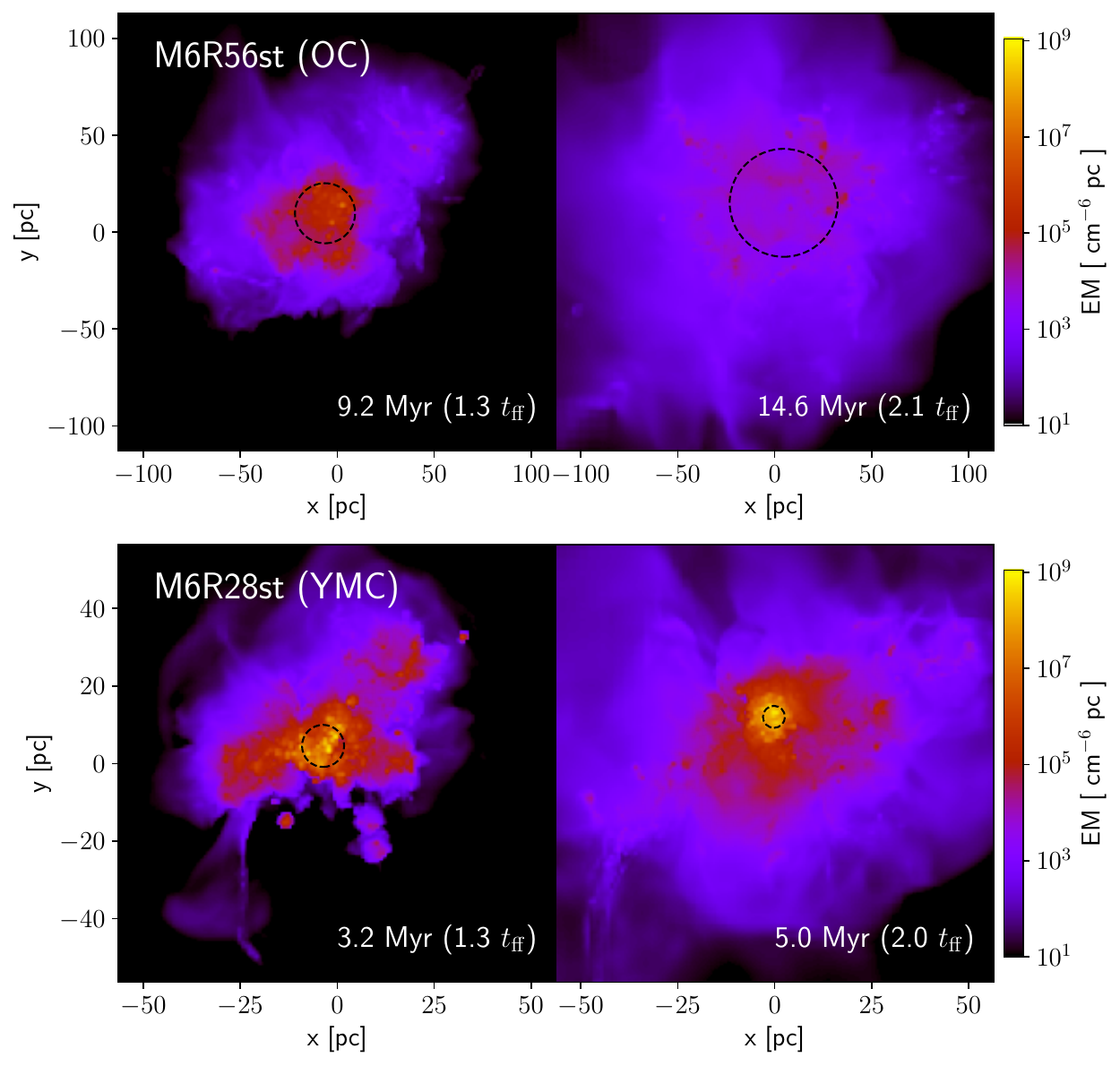}
\caption{EM map for the representative cases of the OC and YMC formation,  M6R56st (top row) and M6R28st (bottom row). In each row, the left and right panels illustrate snapshots at the same epochs as in the second and fourth columns of Figs.~\ref{fig:snap_sigma100m6} and \ref{fig:snap_sigma400m6}. The lines of sight are also the same as these figures. The dashed circle in each panel denotes the half-mass radius of the star cluster measured from its mass centre.
}
\label{fig:EMmap}
\end{figure}


Fig.~\ref{fig:contsed} shows the temporal evolution of the continuum spectra for the representative cases of the normal OC formation (top, case M6R56st) and YMC formation (bottom, case M6R28st), corresponding to the same snapshots as in Figs.~\ref{fig:snap_sigma100m6} and \ref{fig:snap_sigma400m6}.
These panels show the contributions from regions where the local intensity is more than 0.1~\% of the peak values for given frequencies. 
As shown later in Section~\ref{sssec:HIInd}, the size of the emitting regions is $\sim 10$~pc for the YMC-forming case and $\sim$ a few $\times$ 10 - 100~pc for the OC-forming case. The overall shape of the spectrum is common for these cases; one component of the dust thermal emission at $\nu \gtrsim 10^2 - 10^3$~GHz, and the other component of the free-free emission at the lower frequencies.  


Despite the similarity in the spectrum shape, there are striking differences among these cases, for example, in the time evolution. In case M6R56st of the OC formation, the emission gradually declines at all frequencies for $1 \leq t/\tff \leq 2.1$, during which the SFE saturates to $\simeq 0.2$ (Fig.~\ref{fig:mass_history}). The brightness temperature at $\simeq 30$~GHz is always less than 10~K throughout the evolution. This is expected with the standard picture of the \hii bubble expansion; the mean density (and also the column density) of the photoionized gas decreases as the bubble expands.  


In case M6R28st of the YMC formation, in contrast, the emission continues to get stronger for almost the same duration of $1 \leq t/\tff \leq 2$, particularly at $\nu \gtrsim 10^3$~GHz corresponding to the dust thermal emission. The free-free component at the lower frequencies never decreases but rather increases slightly. The brightness temperature at $\simeq 30$~GHz is much higher than for case M6R56st, staying at $\sim 100$~K. These features come from the characteristic evolution of the YMC formation described in Section~\ref{ssec:FY21}. The strong gravity of the forming massive cluster prevents the density of the photoionized gas from decreasing. Recall that the cluster mass continues to increase even after $t = 2~\tff$ (Fig.~\ref{fig:mass_history}).


Another difference is found at the low-frequency end of the free-free continuum spectrum for $\nu \lesssim 10$~GHz. Whereas the spectrum is almost flat for case M6R56st of the OC formation, it decreases with decreasing the frequency for case M6R28st of the YMC formation. This corresponds to the optically thick regime of the free-free emission. Such a feature is known to appear below the turnover frequency
\begin{equation}
\nu_{\rm to} \simeq 16.0~{\rm GHz} 
\left( \frac{{\rm EM}}{10^9 \emunit} \right)^{0.48}
\left( \frac{T_\mathrm{i}}{10^4~\mathrm{K}} \right)^{-0.64} 
\end{equation}
\citep[e.g.][]{Mezger67,Kurtz05,Yang21}, where $T_\mathrm{i}$ is the temperature of the ionized gas and EM represents the emission measure defined as
\begin{equation}
\mathrm{EM} \equiv \int n_e^2 \ ds ,
\end{equation}
for which the integration is performed along the lines of sight. Fig.~\ref{fig:EMmap} shows the evolution of the EM distribution for the cases considered above. In case M6R56st of the OC formation, the peak value of the EM is $\simeq 10^7~\emunit$ in an early stage at $t \simeq 1.3~\tff$, and it decreases by orders of magnitude by the epoch of $\simeq 2.1~\tff$. This is consistent with Fig.~\ref{fig:contsed} (a), where the absorption feature only appears at $\nu \lesssim 1$~GHz in the earliest stage and gradually disappears. 
In contrast, Fig.~\ref{fig:EMmap} shows that for case M6R28st of the YMC formation the EM takes the much higher peak values at $\sim 10^9~\emunit$, which hardly decreases until the latest stage of $t = 2~\tff$. The high EM is due to the trapping effect of the photoionized gas by the gravity of the forming massive cluster. The persistent turnover feature at $\nu \lesssim 10$~GHz in Fig.~\ref{fig:contsed} (b) agrees with the EM evolution presented in Fig.~\ref{fig:EMmap}.

\subsubsection{Density-size diagram of \hii regions}
\label{sssec:HIInd}

\begin{figure}
	\includegraphics[width=0.95\columnwidth]{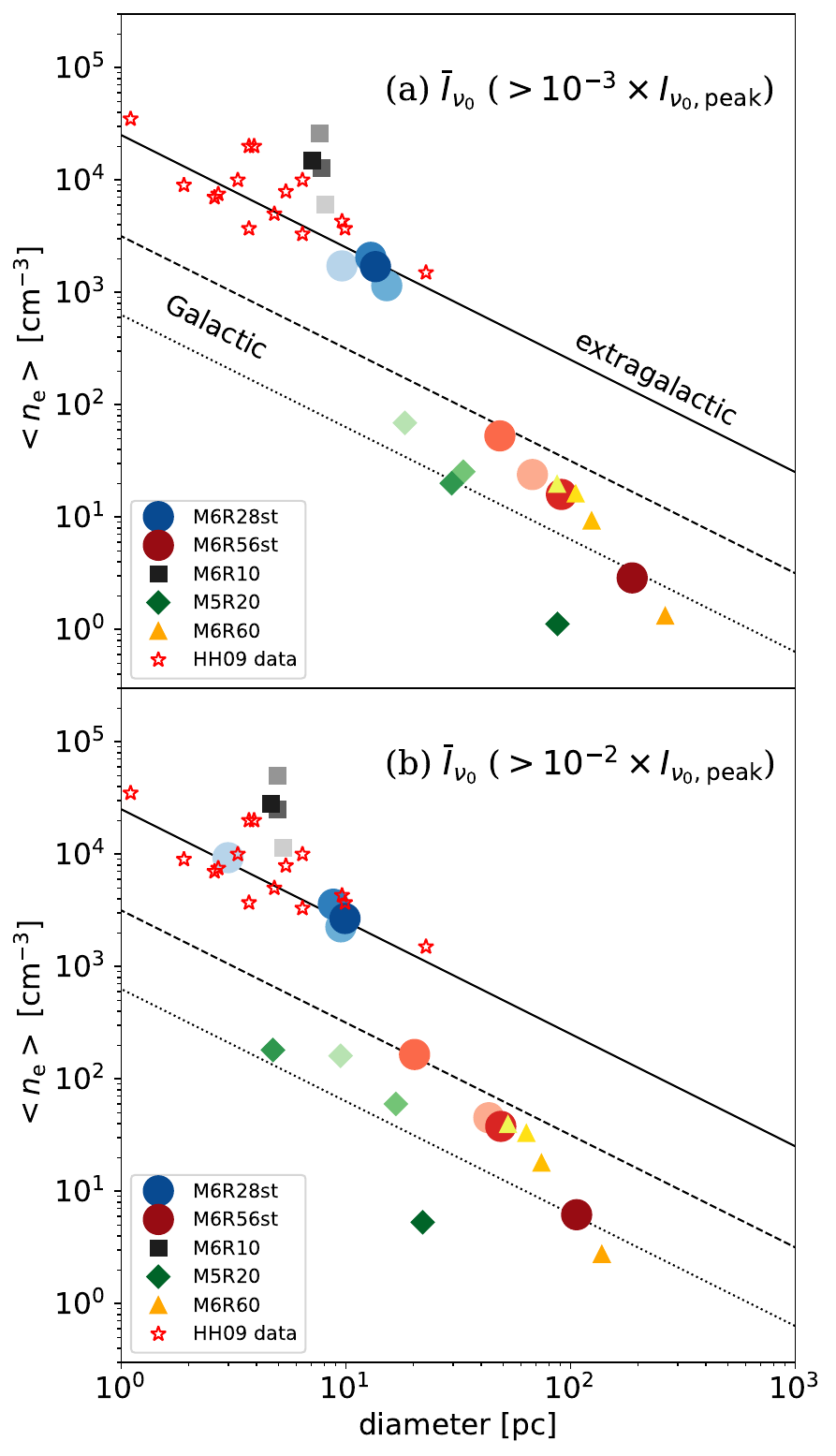}
    \caption{
Mean densities and diameters of \hii regions estimated from emission-measure maps at the frequency of 36 GHz. Panels (a) and (b) represent different threshold intensities for taking a central part, 0.1\% and 1\% of the peak value. 
In both panels, the large blueish and reddish circles represent cases of M6R28st and M6R56st, where the YMC and OC formation eventually occur. The darker colors represent the later evolutionary stages as in Fig.~\ref{fig:contsed}.
We also present additional cases of M6R10, M5R20, and M6R60 with different symbols of the small squares, diamonds, and triangles. M6R10 corresponds to the YMC-formation case, and M5R20 and M6R60 to the OC-formation cases. The darker colors indicate the later stages of $t=1.0$, 1.3, 1.5, and 2.1 $t_\mathrm{ff}$ also for these cases. 
In each panel, the dashed and dotted lines represent the density-size relations of Galactic \hii regions compiled by \citet{Garay1999} and \citet{Kim01}, respectively.
The solid line corresponds to the relation for the extragalactic ultra-dense \hii regions given by \citetalias{Hunt09}. The open red star symbols represent actual samples obtained by radio interferometric observations. 
}
    \label{fig:HIInd}
\end{figure}

One of the important properties provided by radio observations is the density-size relation of \hii regions, which we consider with Fig.~\ref{fig:HIInd}. It has been well established that the Galactic \hii regions obey the density-size relation $\nel \propto D^{-1}$, where $\nel$ is the mean electron number density of an \hii region and $D$ is its diameter \citep[][]{Garay1999, Kim01}. The same correlation continues over \hii regions with sizes that differ by more than four orders of magnitude, from $\sim 0.1$~pc of ultra-compact \hii regions to over 100~pc of giant regions. The black dashed and dotted straight lines in Fig.~\ref{fig:HIInd} represent the best fits to the Galactic samples given by  \citet{Garay1999} and \citet{Kim01}, respectively. Its inclination is somehow shallower than that expected by the simple Str\"omgren-sphere argument assuming a constant emissivity of ionizing photons, $\nel \propto D^{-1.5}$.  


Whether the similar correlation exists for extragalactic \hii regions has been also studied for decades \citep[][]{Kennicut84}. Interestingly, some \hii regions associated with YMCs are known to locate in a distinct region in the density-size diagram. 
Early studies have already suggested the presence of pc-scale \hii regions with high densities above $\nel > 10^3~\cc$ \citep{Kobulnicky99,Beck2002}, which are only found for ultra-compact ($D \lesssim 0.1~\mathrm{pc}$) regions in the Galaxy \citep{Churchwell02}. \citetalias{Hunt09} compile a large sample of such \hii regions in the literature and investigate their statistical properties. The open star symbols in Fig.~\ref{fig:HIInd} represent their radio samples, many of which are \hii regions found in blue compact dwarf galaxies. 
\citetalias{Hunt09} also point out that their density-size relation also approximately follow $\nel \propto D^{-1}$, which corresponds to the solid straight line in Fig.~\ref{fig:HIInd}.


We superpose our simulation data points in Fig.~\ref{fig:HIInd}. To do that, we extract an area where the radio intensity at 36~GHz is higher than some threshold values, 0.1\% and 1\% of the peak value for each snapshot (upper and lower panels). We have confirmed that our results do not change if we shift the frequency to 150~GHz, which is expected by the flat spectrum in this range (Fig.~\ref{fig:contsed}). We compute the effective radius of the area as
\begin{equation}
r = \sqrt{\frac{S}{\pi}} ,
\end{equation}
where $S$ is the area surface, and the mean emission measure $\langle \mathrm{EM} \rangle$ by simply taking the arithmetic average of $\mathrm{EM}$ over cells contained within the area. The mean density $\neav$ is accordingly given by
\begin{equation}
\neav = \sqrt{ \frac{\langle \mathrm{EM} \rangle}{D} },
\end{equation}
where $D$ is the effective diameter $D = 2r$.


As expected with the bimodal evolution for the OC- and YMC-forming cases (Section~\ref{ssec:FY21}), the simulation data points for these cases separately distribute in Fig.~\ref{fig:HIInd}. For case M6R56st of the OC formation (large reddish circles), for instance, the points reside in the relatively lower right portion of each panel. The points go to the lower-right side as time goes on, representing the standard evolution of an \hii bubble expansion where the electron number density gradually decreases with increasing time \citep{Spitzer78}. These points locate near the density-size relation for the Galactic \hii regions. Although the inclination of these points is steeper than $\nel \propto D^{-1}$, reducing the threshold intensity makes them spread widely around the Galactic correlation (dotted and dashed) lines. 


In stark contrast, data points for case M6R28st representing the YMC formation (large bluish circles) are in the relatively upper left portion of each panel, near the extragalactic radio samples by \citetalias{Hunt09}. Furthermore, unlike the OC-forming case described above, these data points hardly move for $t \lesssim 2~\tff$. These remarkable features come from the peculiar evolution of the YMC formation suggested by RHD simulations; an \hii bubble bound by the strong gravity of a newborn cluster postpones the dynamic expansion and retains the high electron density. 
Fig.~\ref{fig:HIInd} suggests that our numerical simulations successfully explain the YMC formation occurring in nearby starburst galaxies. 


In addition to the two cases above representing the OC and YMC formation, we also perform the same post-process calculations for other cases studied in \citetalias{FY21} (Table~\ref{tab:simpar}). As indicated by small filled symbols representing these cases, the distribution of the simulation data is bimodal, reflecting the bimodal evolution of the OC- and YMC-forming cases. 
We conclude that the distinct distribution of extragalactic \hii regions on the density-size diagram is the signature of the YMC formation, i.e., that of the long-lasting dense \hii regions bounded by gravity.

\subsubsection{Dust thermal emission}
\label{sssec:dust}

\begin{figure*}
	\includegraphics[width=1.95\columnwidth]{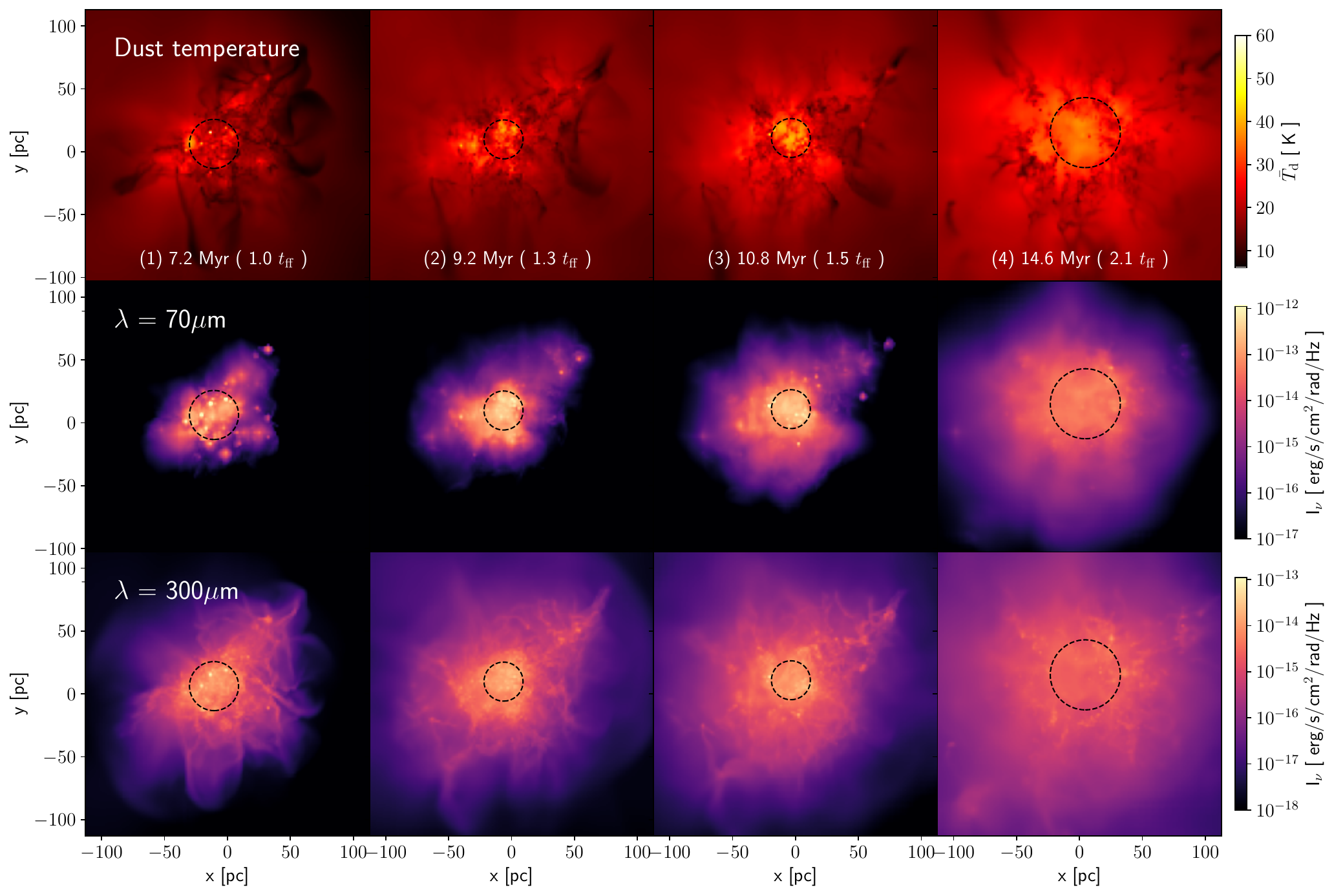}
\caption{Dust continuum emission maps for the representative case of the OC formation, M6R56st. The top, middle, and bottom rows show the time sequence of the distributions of the column-density weighted dust temperature, and emission intensities at the wavelengths $\lambda = 70$ and 300\micron. The columns of panels show the snapshots at the same epochs as in Fig.~\ref{fig:snap_sigma100m6}. In each panel, the dashed circle denotes the half-mass radius of the star cluster measured from its mass centre.
}
\label{fig:DImap_sigma100m6}
\end{figure*}
\begin{figure*}
\includegraphics[width=1.95\columnwidth]{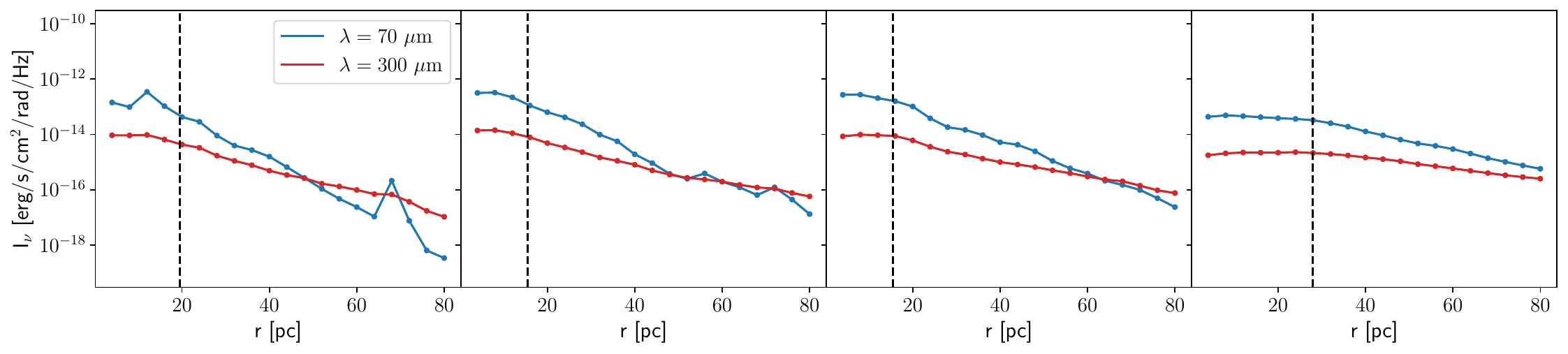}
\caption{Mean radial distributions of the dust continuum emission intensities for the representative case of the OC formation, M6R56st. In each panel, the points represent the intensities averaged over rings centered on the cluster mass centre. The blue and red colors correspond to the different wavelengths of $\lambda = 70$ and 300\micron. The panels show the snapshots at the same epochs as in Figs.~\ref{fig:snap_sigma100m6} and \ref{fig:DImap_sigma100m6}. The vertical dashed line in each panel represents the half-mass radius of the star cluster.   
}
\label{fig:DImap_radial_sigma100m6}
\end{figure*}
\begin{figure*}
	\includegraphics[width=1.95\columnwidth]{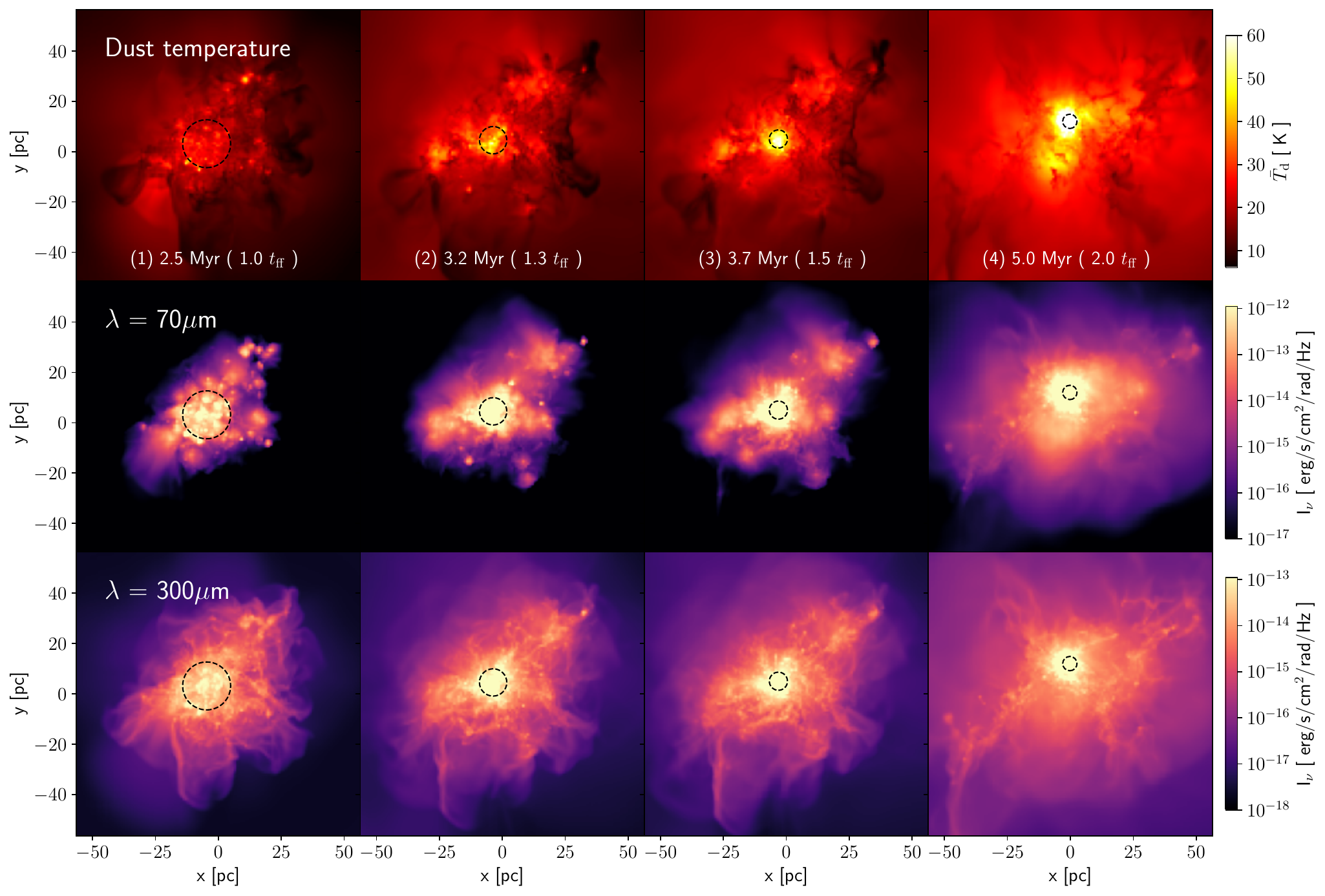}
\caption{The same as in Fig.~\ref{fig:DImap_sigma100m6} but for the representative case of the YMC formation, M6R28st. The columns of panels show the snapshots at the same epochs as in Fig.~\ref{fig:snap_sigma400m6}. }
\label{fig:DImap_sigma400m6}
\end{figure*}
\begin{figure*}
\includegraphics[width=1.95\columnwidth]{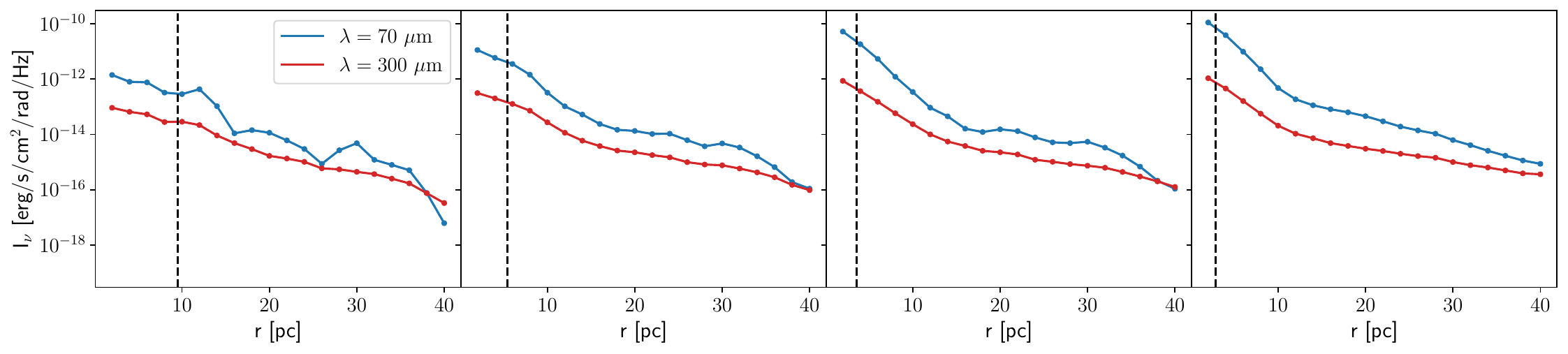}
\caption{The same as Fig.~\ref{fig:DImap_radial_sigma100m6} but for the representative case of the YMC formation, M6R28st. The panels show the snapshots at the same epochs as in Figs.~\ref{fig:snap_sigma400m6} and \ref{fig:DImap_sigma400m6}.   }
\label{fig:DImap_radial_sigma400m6}
\end{figure*}

As already suggested by Fig.~\ref{fig:contsed}, the dust thermal emission at far-IR wavelengths also shows the characteristic features of the OC and YMC formation. 
Fig.~\ref{fig:DImap_sigma100m6} shows the evolution of the column-density-weighted dust temperature $\Td$ and continuum emission maps at $\lambda = 70~\micron$ and $300~\micron$ for the representative case of the OC formation, M6R56st. The dust temperature is an indicator of the strength of the local stellar radiation field, the heating source for grains. In fact, the dust temperature is highest near the cluster centre at each snapshot. The peak value gradually rises for $\tff \lesssim t \lesssim 1.5~\tff$, during which the cluster mass increases (Fig.~\ref{fig:mass_history}). However, it decreases for $1.5 \tff \lesssim t \lesssim 2~\tff$, because the cluster spatially expands only with a slight increase in mass (Fig.~\ref{fig:snap_sigma100m6}). The dust temperature at $\sim 100$~pc away from the cluster centre monotonically increases as the cluster expands and the cloud disperses instead. 
For a given snapshot, the emission at $\lambda = 70~\micron$ is more centrally concentrated than at $300~\micron$ because it traces the distribution of warmer grains. The radial extent of the emission at these wavelengths gradually expands with time as the cluster expands. The emission is always strongest near the cluster centre, but its peak values substantially decrease for $1.5 \tff \lesssim t \lesssim 2~\tff$. This is because the \hii bubble expansion reduces the column density of the gas and dust, in addition to the decrease of $\Td$. Note that the dust-to-gas mass ratio is homogeneous within the computational domain in our simulations as in \citetalias{FY21}. Fig.~\ref{fig:DImap_radial_sigma100m6} also clearly shows such evolution, presenting mean radial distributions of the dust continuum emission intensities at each snapshot.


Figs.~\ref{fig:DImap_sigma400m6} and \ref{fig:DImap_radial_sigma400m6} are the same as Figs.~\ref{fig:DImap_sigma100m6} and \ref{fig:DImap_radial_sigma100m6} but for the representative case of the YMC formation, M6R28st. 
Reflecting the different evolutions outlined in Section~\ref{ssec:FY21} among these cases, these figures also show the different features from those described above. For instance, Figs.~\ref{fig:DImap_sigma400m6} shows that the dust temperature $\Td$ near the cluster centre is much higher than in case M6R56st for $t \gtrsim 1.3~\tff$, because the cluster mass continues to increase until the final snapshot. Since the cluster does not expand, unlike case M6R56st, the stellar radiation field at the cluster centre gets stronger and stronger while the cluster grows in mass.  
The evolution of the emission maps at $\lambda = 70$ and $300~\micron$ also differs from the above. Most remarkably, the maps in the later stages of $t \gtrsim 1.5~\tff$ show strong and centrally concentrated emission at both wavelengths. 
In addition to the high $\Td$, this is also due to the difference in the gas density structure. Since the cluster's gravity is strong enough to even trap the photoionized gas, there remains the dense gas in the vicinity of the cluster. The column densities of the gas and dust hardly decrease for a long time. Fig.~\ref{fig:DImap_radial_sigma400m6} suggests that for $t \gtrsim 1.5~\tff$ the peak intensities in case M6R28st are higher than those in case M6R56st by a few orders of magnitude. At the final snapshot at $t \simeq 2~\tff$, there are sharp cusps within the central $\sim$ 10~pc, which is in stark contrast to the flat core-like distributions seen for case M6R56st (c.f. Fig.~\ref{fig:DImap_radial_sigma100m6}).

\section{Discussions}
\label{sec:discussion}

\subsection{Alternative signatures: line emission and absorption}

Our focus has been on the continuum spectra to derive characteristic signatures of the YMC formation, but this should not be the only available observational clue. Alternative ways include line emission against the continuum spectra, which is to be further studied. 
Various lines spanning from radio to infrared wavelengths are expected to emerge from different gas phases surrounding YMCs during their formation stage. For instance, infrared emission lines from a photodissociated region (PDR) are promising candidates \citep[e.g.][and references therein]{Berne2022}. Similar to the dust thermal emission studied in Section~\ref{sssec:dust}, these line emission maps will exhibit distinct features that differentiate between OC- and YMC-forming cases. This is one of our next projects in progress. 


An advantage of the line features is allowing us to infer the gas kinematics around the forming YMC. Our preparatory analyses of the simulation data show that, for the YMC-forming cases, the cluster's strong gravity causes the infall motion of the surrounding gas toward the centre of an \hii region. Such a signature of the infall motion has been really reported on Galactic ultra-compact \hii regions \citep{Sollins05,Beltran2006}. These studies make use of the absorption feature of foreground molecular (NH$_3$) inversion lines against the free-free continuum emission \citep{Ho1983}.  
According to our simulations, the \hii region that emerges in the YMC formation is much larger than the typical ultra-compact \hii regions by two orders of magnitude. Nonetheless, they might show similar observational signatures suggesting the infall motion in future observations. 


Throughout this paper, we have mainly considered the YMC formation in the nearby universe. As mentioned in Section~\ref{sec:intro}, however, YMC formation may play a more significant role in star formation in the early universe. Recent observations with the JWST have found more galaxies at $z > 10$ than expected \citep[e.g.][]{Naidu22,Harikane23}, which prompts intensive studies and discussions \citep[e.g.][]{Boylan-Kolchin22,Yajima22,Lovell23,Dekel23}. One possible explanation is that YMC formation, which achieves a high SFE, is the dominant mode of star formation \citep{Inayoshi22}. In observations of distant galaxies, restframe far-IR line emissions such as [\oi]~63$\micron$ and [\cii]~158$\micron$ coming from PDRs are often used because they redshift into the ALMA bands. In the same vein, [\oiii] 88$\micron$ from \hii regions is a well-known tracer \citep{Inoue16,Hashimoto18,Witstok22}. All of these line emissions can be predicted for both OC- and YMC-forming cases using our simulation data. Although individual \hii regions cannot be resolved in distant galaxies, if YMC formation is dominant for the overall star formation in galaxies at $z>10$, it will be essential to comprehensively examine these line emissions. However, in such extreme environments, other effects that we have not considered in this paper, such as very low metallicity or a different IMF \citep{Zackrisson11,Chon21,Chon22,Fukushima23}, may also come into play. It is also interesting to study emission line diagnostics for YMC formation, taking into account these effects.

\subsection{Possible signatures in an earlier evolutionary stage}


While our simulation data predict observational signatures consistent with the radio observations (e.g. Fig.~\ref{fig:HIInd}), the simulation runs start from the idealized initial conditions. 
Such situations of isolated molecular clouds are not necessarily realized in the YMC formation. Although the actual triggers of the YMC formation are still uncertain, cloud-cloud collisions or large-scale converging flows have been proposed as a possible process \citep[e.g.][]{Maeda21,Sameie22,Dobbs22}. 
Such realistic initial conditions for YMC formation should be provided by the general framework of galactic-scale star formation \citep[e.g.][]{Inutsuka+15}.
Our lack of knowledge about the realistic initial conditions of YMC formation prevents us from predicting observational signatures during the early stages of YMC formation.
\citet{Rico-Villas20} and \citet{Rico-Villas22} conduct diagnostic analyses with many molecular lines detected toward ``proto-superstar clusters'' candidates in NGC 253, and consider possible evolutionary sequences in the YMC formation. They suggest that there is the ``super hot core'' stage in the YMC formation before the appearance of an \hii region, similar to the hot core stage in the individual high-mass star formation \citep{Garay1999,Hoare07}. 
To improve the accuracy of our predictions and better compare them with such observations, we need to update our simulations to reflect more realistic initial conditions.
Nevertheless, we expect that our results on the observational signatures of the gravitationally bound \hii regions should not depend on details of the initial conditions, as long as the key process enabling continuous star formation against the stellar UV feedback is accurately captured.
This also remains to be examined in future studies.

\subsection{Dust grains within \hii regions}

RHD simulations by \citetalias{FY21} make some basic assumptions about the dust grains within the \hii regions. They assume that the gas and dust are dynamically coupled, resulting in a uniform dust-to-gas mass ratio. They also assume that the grain size distribution and dust opacity are uniform. The dust sublimation is not taken into account for simplicity. Here, we consider the potential uncertainties that arise from these assumptions.


Theoretical studies have suggested that dust particles can move independently of the gas, particularly in the vicinity of luminous stars where the grains are exposed to a strong radiation force \citep[e.g.][]{Gail79,Draine11b,Fukushima18}. \citet{Akimkin15,Akimkin17} investigate this effect during the expansion of \hii regions by performing 1D RHD simulations. They show that the dust-gas decoupling occurs within an \hii region, and the degree of relative motion depends on the size and charge of the grains. This implies that the grain size distribution and the dust-to-gas mass ratio can be variable within the \hii region. Consequently, the dust continuum emission maps discussed in Section~\ref{sssec:dust} will be altered. \citet{Ishiki18} point out that, within \hii regions excited by massive clusters, the segregation between small and large grains is alleviated by the efficient Coulomb drag force acting on highly charged grains. This may be the case for the formation of YMCs, which will be explored in future studies. 


Note that \citetalias{FY21} take into account the absorption of stellar ionizing radiation by dust grains within \hii regions, which reduces the sizes of \hii regions \citep[e.g.][]{Petrosian72, Spitzer78, inoue01, Arthur04}. The resulting radiation force exerted on the dust grains is also incorporated under complete dynamical coupling with the gas. 
It has been proposed that the strong radiation force from high-mass stars creates central dust cavities within \hii regions, often observed in \citep{Inoue02,Draine11b,Kim16}.
In the case of M6R28st, which forms a YMC, strong feedback caused by the radiation force eventually becomes effective in disrupting the accretion flow into the cluster and terminating star formation. This occurs for $t \gtrsim 3.5~t_\text{ff}$, after the last snapshot considered above.


Regarding the dust sublimation, recall that the dust temperature is determined considering the heating/cooling balance in \citetalias{FY21}. In any case of their simulation runs, the dust temperatures within \hii regions are much lower than $1000$~K, the temperature above which sublimation typically occurs. This implies that dust sublimation does not significantly affect global dust distribution within the pc-scale or \hii regions. Although a dust sublimation front should be present near an individual massive star, it is not resolved in \citetalias{FY21}, which may lead to an underestimation of the strength of the radiation-force feedback \citep{Krumholz18b}. \citetalias{FY21} provide an appendix to check the numerical convergence with increasing resolution.

\section{Summary and Conclusions}
\label{sec:conclusion}

We have studied possible observational signatures of YMCs in their formation stage, i.e., the evolutionary stage where the cluster mass grows through accretion from the surrounding medium. In particular, we have considered the continuum spectra from $\sim$ 10~pc-scale dense \hii regions created around the accreting YMC, as suggested by recent RHD simulations \citepalias{FY21}. 
We have performed post-process radiative transfer calculations using the simulation snapshots to derive the spectra from radio to far-IR frequencies. For comparisons, we have also performed the same post-process calculations for the cases where normal OCs eventually appear instead of YMCs. Our findings are summarized as follows.


For both simulation runs that represent OC and YMC formation, the continuum spectrum commonly consists of two components: one dominated by the thermal dust emission at $\nu \gtrsim 10^2 - 10^3$~GHz, and the other dominated by the free-free emission at the lower frequencies. However, there are remarkable differences between these cases, as illustrated in Fig.~\ref{fig:contsed}. The spectrum for the normal OC-forming case (M6R56st) represents the standard picture of an \hii bubble expansion, where the electron density gradually decreases as the bubble expands.
 The intensities also decline from the radio to infrared for $\tff \lesssim t \lesssim 2 \tff$, during which the radiative feedback by the expanding bubble destroys the natal cloud and quenches the star formation. For the YMC-forming case (M6R28st), in contrast, the intensities gradually rise for the same time interval normalized by $\tff$ and are always much stronger than those for the OC-forming case. Moreover, there is a turnover feature below $\sim 10$~GHz resulting from the large emission measure. These all come from the peculiar evolution in the YMC-forming case, i.e., the rapid star formation under weak radiative feedback, which leads to the formation of a dense \hii region that remains trapped by the cluster's gravitational field.


Previous radio observations provide density-size relations of Galactic and extragalactic \hii regions, which have been investigated to infer the variety of high-mass star- and cluster-forming environments \citep{Garay1999,Kim01}. Remarkably, it has been pointed out that some extragalactic \hii regions distribute separately from Galactic \hii regions \citepalias{Hunt09}. They are large ($\sim 10$~pc), dense ($\sim 10^4~\cc$), and associated with YMCs. 
We have superposed our simulation data points on the density-size diagram based on the synthetic continuum emission maps at 36~GHz (Fig.~\ref{fig:HIInd}). Reflecting the qualitatively different evolution between the OC- and YMC-forming cases, the corresponding data points distinctly distribute on the diagram. For the normal OC-forming cases, the simulation points are situated close to the Galactic density-size relationships. As the evolution progresses to later stages, the points shift to the less-dense and larger-size portion, in accordance with the standard expansion law of \hii bubbles. The points for the YMC-forming cases, in contrast, scatter separately from the OC-forming cases and near the observational data of extragalactic radio samples. These points hardly move on the diagram regardless of $t/\tff$, reflecting that the cluster's gravity traps an \hii bubble and prevents it from dynamically expanding for a while. We propose that previous radio observations have already captured the signatures of the YMC formation suggested by recent RHD simulations. 


We suggest that similar trends to the above should also be found in the far-IR continuum dust thermal emission maps (Figs.~\ref{fig:DImap_sigma100m6}--\ref{fig:DImap_radial_sigma400m6}). We have investigated the evolution of the maps at $\lambda = 70~\micron$ and $300~\micron$.
For the normal OC-forming case, the radial emission distribution gradually broadens with time. As a result, the emission peak associated with the cluster centre shows a decrease in intensity. A flat, core-like distribution persists within the half-mass radius of the cluster. This can be attributed to the dynamic expansion of the \hii bubble, as well as to the outward expansion of the star cluster induced by the dispersal of the surrounding molecular cloud.
The YMC-forming case, in contrast, shows that the central emission peak does not weaken but rather becomes stronger over time.
The radial emission distribution evolves to show a remarkably high degree of central concentration, characterized by a sharp peak at the centre of the cluster. This is again due to the emergence of a dense \hii bubble gravitationally bound by the nascent YMC, which continues to accrete mass.


Our study suggests that the peculiar YMC formation process found in recent RHD simulations, i.e., the formation of gravitationally-bound dense \hii regions, should indicate observational signatures. We conclude that such signatures have been captured by extragalactic radio observations. 
As discussed in Section~\ref{sec:discussion}, our work can be easily stretched to further studies that link simulations and upcoming observations. Particularly, considering observational signatures of emission and absorption lines associated with the YMC formation is of great importance, and it is also a target of our next work.


\section*{Acknowledgements}

The authors sincerely appreciate Jeong-Gyu Kim, Hiroyuki Hirashita, Eric Keto, Rolf Kuiper, Yurina Nakazato, Akino Inoue, and Shu-ichiro Inutsuka for the comments and discussions. The numerical simulations were carried out on XC50  {\tt Aterui II} at the Center for Computational Astrophysics (CfCA) of the National Astronomical Observatory of Japan. This research could never be accomplished without the support by Grants-in-Aid for Scientific Research (TH:19H01934, 21H00041, HF:23K13139, KEIT: 19K14760, 21H00058, HY:21H04489) from the Japan Society for the Promotion of Science and JST FOREST Program, Grant Number JPMJFR202Z (HY).

\section*{Data Availability}

The data underlying this article will be shared on reasonable request to the corresponding author.



\bibliographystyle{mnras}
\bibliography{ref} 




%
%


\bsp	
\label{lastpage}
\end{document}